\documentclass[twocolumn,aps,prl,reprint,citeautoscript,superscriptaddress,showkeys,showpacs]{revtex4-1}
\usepackage[latin9]{inputenc}
\usepackage{amsmath}
\usepackage{amssymb}
\usepackage{graphicx}

\makeatletter
\usepackage{xr}
\usepackage{amsfonts}
\usepackage{amsxtra}\usepackage{textcomp}\usepackage{grffile}
\usepackage{bbm}
\usepackage{bm}
\usepackage{color}
\usepackage{LatexCommands}

\makeatother

\begin{document}

\title{Impact of damping on superconducting gap oscillations induced by
intense Terahertz pulses}

\author{Tianbai Cui}

\affiliation{School of Physics and Astronomy, University of Minnesota, Minneapolis,
Minnesota 55455, USA}

\author{Xu Yang}

\affiliation{Department of Physics and Astronomy, Iowa State University, Ames,
Iowa 50011, USA}

\affiliation{Ames Laboratory, U.S. DOE, Iowa State University, Ames, Iowa 50011,
USA}

\author{Chirag Vaswani}

\affiliation{Department of Physics and Astronomy, Iowa State University, Ames,
Iowa 50011, USA}

\affiliation{Ames Laboratory, U.S. DOE, Iowa State University, Ames, Iowa 50011,
USA}

\author{Jigang Wang}

\affiliation{Department of Physics and Astronomy, Iowa State University, Ames,
Iowa 50011, USA}

\affiliation{Ames Laboratory, U.S. DOE, Iowa State University, Ames, Iowa 50011,
USA}

\author{Rafael M. Fernandes}

\affiliation{School of Physics and Astronomy, University of Minnesota, Minneapolis,
Minnesota 55455, USA}

\author{Peter P. Orth}

\affiliation{Department of Physics and Astronomy, Iowa State University, Ames,
Iowa 50011, USA}

\affiliation{Ames Laboratory, U.S. DOE, Iowa State University, Ames, Iowa 50011,
USA}
\begin{abstract}
We investigate the interplay between gap oscillations and damping
in the dynamics of superconductors taken out of equilibrium by strong
optical pulses with sub-gap Terahertz frequencies. A semi-phenomenological
formalism is developed to include the damping within the electronic
subsystem that arises from effects beyond BCS, such as interactions
between Bogoliubov quasiparticles and decay of the Higgs mode. Such
processes are conveniently expressed as $T_{1}$ and $T_{2}$ times
in the standard pseudospin language for superconductors. Comparing
with data on NbN that we report here, we argue that the superconducting
dynamics in the picosecond time scale, after the pump is turned off,
is governed by the $T_{2}$ process. 
\end{abstract}
\maketitle

\emph{Introduction.\textendash{}} The coherent control of non-equilibrium
states of interacting quantum matter promises far-reaching capabilities
by turning on (or off) desired electronic material properties. A particular
focus in this field has been the manipulation of superconductivity
by non-equilibrium probes. While earlier works showed that microwave
pulses could be used to enhance the superconducting transition temperature
$T_{c}$ of thin superconducting films~\cite{Eliashberg-1960,Pals-PhysRep-1989},
recent advances in ultrafast pump-and-probe techniques opened the
possibility of investigating superconductivity in the pico- and femto-second
timescales by coherent light pulses~\cite{PhysRevLett.111.057002,MatsunagaShimano-Science-HiggsMode-2014}.
Such coherent pulses have been employed to manipulate the electronic
and lattice properties of quantum materials, resulting in transient
behaviors that are consistent with the onset of non-equilibrium superconductivity
above $T_{c}$~\cite{Mankowsky-Nature-2014,Fausti14012011,2015arXiv150504529M}.
Alternatively, coherent pulses have also been employed to assess the
coherent dynamics of the superconducting state~\cite{PhysRevLett.109.187002, PhysRevLett.111.057002, MatsunagaShimano-Science-HiggsMode-2014, 0295-5075-101-1-17002, PhysRevB.91.214505, PhysRevB.93.094509, PhysRevLett.103.075301, Krull2016}.

To maintain coherence and avoid excess heating, it is advantageous
to apply pulses at energies below twice the superconducting gap $2\Delta$,
where quasi-particle (Bogoliubov) excitations are absent. As the superconducting
gap energy scale lies in the Terahertz (THz) regime, this requires
the application of intense and coherent sub-gap THz light pulses~\cite{Kampfrath-NaturePhotonics-2013}.
In Ref.~\cite{PhysRevLett.111.057002}, a monocycle intense THz pulse
was applied to a thin film of the conventional $s$-wave superconductor
NbN, reporting coherent oscillations of the superconducting gap with
frequency $2\Delta$.

Such oscillations arise naturally from the solution of the time-dependent
BCS (Bardeen-Cooper-Schrieffer) equation~\cite{Volkov1974, Yuzbashyan-BCS-Exact-Solution, Warner_Leggett, PhysRevLett.93.160401, PhysRevB.72.220503, Sentef-trARPES, PhysRevB.90.014515, Higgs_Spectroscopy}, which can be conveniently recast in terms of Anderson pseudospins~\cite{Anderson-RPA-BCS}
precessing around a pseudo magnetic field that is changed by the optical
pulse. While this coherent evolution describes qualitatively well
the behavior of the system in a restricted time window, there is also
damping present in the system, which is absent in this BCS approach.

Here, we develop a semi-phenomenological model that captures not only
the coherent evolution of the gap function, in the picosecond time
scale, but also damping effects in the time scale of tens to hundreds
of picoseconds. Since this time scale precedes the thermalization
with the lattice, the relevant relaxation processes arise within the
electronic subsystem from effects not captured by BCS. These include
interactions between Bogoliubov quasiparticles and the coupling between
the Higgs (amplitude) mode and the continuum. In the pseudospin notation,
we identify two types of relaxation process: the longitudinal relaxation
$T_{1}$, corresponding to relaxation of quasiparticles, and the transverse
relaxation $T_{2}$, corresponding to relaxation of the gap.

We apply this formalism to elucidate the dynamics of superconducting
NbN, which was measured at very low temperatures using intense THz
fields with sub-gap spectra. Our data reveals the gap oscillating
at a frequency corresponding to twice the pump frequency. When the
pump is turned off, however, the gap oscillations quickly disappear,
and the amplitude of the gap continues to be suppressed. Such a behavior
is at odds with the nonequilibrium dynamics given by the time-dependent
BCS equation, where the gap displays coherent oscillation with very
slow collisionless relaxation~\cite{Volkov1974}. We show instead
that this behavior is well captured by our semi-phenomenological model,
and arises from a dominant $T_{2}$ relaxation process whose time
scale is of the same order as the duration of the pump.

\begin{figure}
\begin{centering}
\includegraphics[width=0.9\columnwidth]{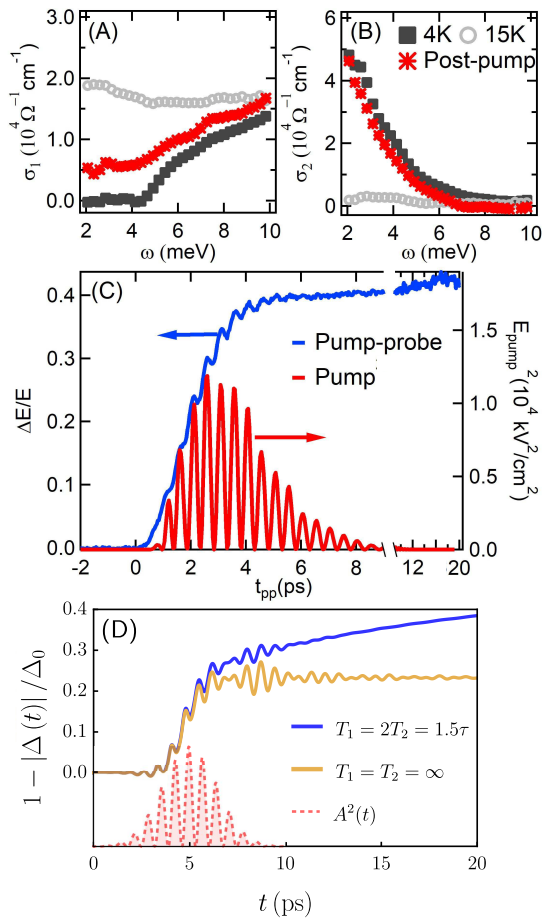} 
\par\end{centering}
\caption{THz pump-probe specroscopy of NbN. (A) and (B) depict the real and
imaginary parts of the optical conductivity, respectively. Gray curves
show equilibrium results at $T=4$ K (below $T_{c}$) and $T=15$
K (above $T_{c}$), whereas the red curve is taken $t_{\mathrm{pp}}=10$
ps after the THz pump. (C) Relative pump-induced change of the transmitted
probe field strength $\Delta E/E$ (blue curve). The value of $t_{\mathrm{gate}}$
is chosen as to be sensitive to changes in the transmittance around
$4$ meV. The red curve shows the pump profile. (D) Theoretical results
for the gap evolution without (yellow line) and with (blue line) damping.
The time scales $T_{1}$ and $T_{2}$ refer to the relaxation processes
explained in the main text. \label{fig_exp}}
\end{figure}

\emph{Experimental results.\textendash{}} The data was acquired using
an intense THz pump, weak THz probe ultrafast spectroscopy setup.
A Ti-Sapphire amplifier was used to generate pulses of energy $3$
mJ, duration $40$ fs, 1 KHz repetition rate, and $800$ nm center
wavelength. The pulses were split into three paths: pump, probe and
sampling. The intense THz pump pulses were generated by the tilted-pulse-front
phase matching through a 1.3\% MgO doped LiNbO$_{3}$ crystal. The
weak THz probe pulses, generated by optical rectification, were detected
by free space electro-optic sampling through a 1mm thick (110) ZnTe
crystal. Pump and probe THz pulses with orthogonal polarizations were
combined by a wire grid polarizer in a collinear geometry and focused
on the sample at normal incidence. The pump was blocked by another
wire grid polarizer and the probe electric field $E$ was sampled
by a $800$ nm pulse. The peak $E$ field of the narrow-band $1$
THz pump was observed to be as large as $109$ kV/cm.

The sample studied here was a $120$ nm NbN film grown on (100)-oriented
MgO single crystalline substrates via pulsed laser deposition, as
previously reported in Ref.~\cite{PhysRevB.93.180511}. The equilibrium
and non-equilibrium optical conductivity were extracted from the complex
transmission using a scanning gate pulse delay $t_{\mathrm{gate}}$.
The ultrafast dynamics was extracted by scanning the optical delay
$t_{\mathrm{pp}}$ between the pump and the probe.

Figs. \ref{fig_exp} (A)-(B) show the behavior of the real and imaginary
parts of the optical conductivity, $\sigma_{1}\left(\omega\right)$
and $\sigma_{2}\left(\omega\right)$, respectively. In equilibrium
(gray cruves), the onset of superconductivity below $T_{c}\approx13.4$
K is signaled by the opening of a gap $2\Delta\approx4.2$ meV in
$\sigma_{1}\left(\omega\right)$, and by a $1/\omega$ dependence
of $\sigma_{2}\left(\omega\right)$ at low frequencies. The post-pump
state (red curve) exhibits larger values of $\sigma_{1}\left(\omega\right)$
within the $2\Delta$ range, and slightly reduced values of $\sigma_{2}\left(\omega\right)$,
presumably due to the THz-induced quench of the SC condensate~\cite{PhysRevLett.109.187002}.

To extract the ultrafast dynamics of the gap function, we measure
the change in the transmitted field $\Delta E/E$, which was shown
in Ref.~\cite{MatsunagaShimano-Science-HiggsMode-2014} to faithfully
reflect the transient behavior of $\Delta\left(t\right)$. Fig. \ref{fig_exp}(C)
shows the ultrafast time evolution of $\Delta E/E\propto1-|\Delta(t)|/\Delta_{0}$,
with $\Delta_{0}\equiv\Delta(t=0)$, well inside the superconducting
state (blue curve, at $T=4$ K), superimposed with the applied pump
pulse (red curve). Interestingly, we find oscillations on $\Delta(t)$
only while the pump pulse is on. After it is turned off, the oscillations
disappear quickly, but $\Delta(t)$ continues to decrease on the time
scale of tens of picoseconds. The Fourier decomposition of $\Delta E/E$
(not shown) indeed demonstrates that the gap oscillations do not scale
with the gap function, unlike reported for shorter monocycle pulses~\cite{PhysRevLett.111.057002},
but instead correspond to twice the pump frequency~\cite{MatsunagaShimano-Science-HiggsMode-2014}.

\emph{Theoretical modeling and analysis.\textendash{}} To model and
elucidate these experimental results, we need to consider relaxation
processes beyond the standard coherent time evolution predicted in
BCS theory. Within BCS, the quench dynamics of $\Delta(t)$ can display
three different behaviors \cite{Volkov1974,PhysRevLett.96.097005,PhysRevLett.96.230403,PhysRevLett.96.230404}:
(i) overdamped decay of $\Delta(t)\rightarrow0$ (phase I); (ii) underdamped
oscillations with frequency $2\Delta_{\infty}$ that decay algebraically
($\propto t^{-1/2}$) towards a finite asymptotic value $\Delta(t)\rightarrow\Delta_{\infty}$
(phase II); and (iii) persistent undamped oscillations (phase III).

In contrast to these predictions, our experimental observation is
that although the gap oscillations are rapidly damped out, the gap
remains finite after the pump pulse is off (see Fig.~\ref{fig_exp}(C)).
Moreover, it continues to show a slow decay between $10$~ps and
$20$~ps, a behavior that presumably persists into the time scale
of hundreds of picoseconds. The gap eventually returns to its initial
equilibrium value on even longer nanosecond time scales \emph{via}
equilibration with phonons. This regime is not discussed in this paper.

To explain this discrepancy, one must include damping within the electronic
subsystem. Before discussing possible microscopic mechanisms for damping,
we employ a phenomenological approach that is best expressed within
the pseudospin description of the BCS model. The standard BCS Hamiltonian
is given by:
\begin{equation}
H_{\text{BCS}}=\sum_{\bfk,\sigma}\xi_{\bfk+e_{0}\bfaa}c_{\bfk,\sigma}^{\dag}c_{\bfk,\sigma}-\sum_{\bfk}\bigl(\Delta c_{\bfk,\uparrow}^{\dag}c_{-\bfk,\downarrow}^{\dag}+\text{h.c.}\bigl)+\frac{|\Delta|^{2}}{V_{0}}\label{BCS}
\end{equation}

Here we consider the square-lattice dispersion $\varepsilon_{\bfk}=-2J(\cos k_{x}+\cos k_{y})$,
and $\xi_{\bfk}=\varepsilon_{\bfk}-\mu$, with chemical potential
$\mu=-1.18J$ corresponding to quarter-filling, and electron charge
$e_{0}$. The superconducting order parameter obeys the self-consistent
equation $\Delta=-V_{0}\sum_{\bfk}\av{c_{-\bfk,\downarrow}c_{\bfk,\uparrow}}$,
where $V_{0}<0$ denotes an attractive interaction. For the calculations
in this paper, we set $V_{0}=-3J$ and the Debye frequency $\omega_{D}=J/2$,
yielding $\Delta_{0}=0.08J$ and $T_{c}=0.048J$.

\begin{figure}
\begin{centering}
\includegraphics[width=1\columnwidth]{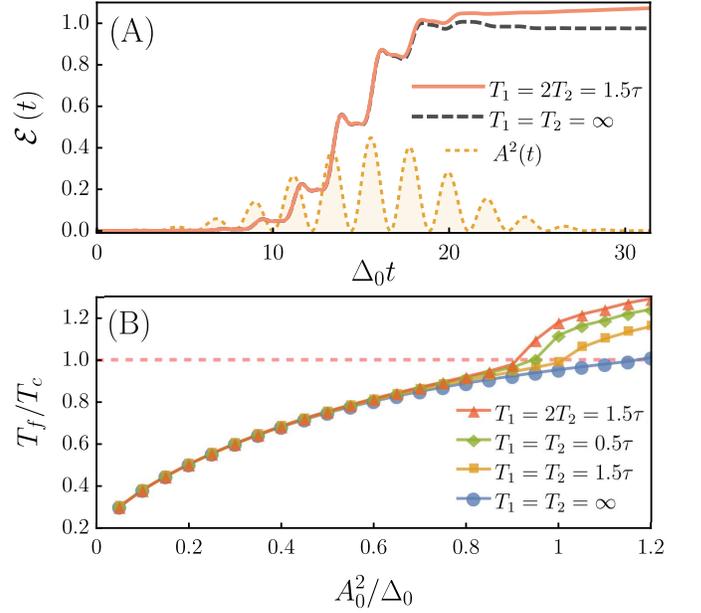} 
\par\end{centering}
\caption{(A) The time evolution of the internal energy of the electronic subsystem
for $T_{1}=T_{2}=\infty$ (dashed) and $T_{1}=2T_{2}=1.5\tau$ (red)
arising from the energy deposited by the pump. The energy is normalized
by $N_{f}\Delta_{0}^{2}$, where $N_{f}$ is the density of states at the Fermi level. (B) The effective temperature after the
pump is turned off, $T_{f}\equiv T^{*}\left(\tau\right)$, normalized
by $T_{c}$, as a function of the pump intensity for various $T_{1}$
and $T_{2}$. For finite $T_{1,2}$, the system will relax to the normal state once $T_{f} > T_{c}$, which leads to an increased energy absorption (as indicated by the change of slope of $T_{f}$ when crossing the red dashed line).}
\label{fig_theory2} 
\end{figure}

\begin{figure}
\begin{centering}
\includegraphics[width=1\columnwidth]{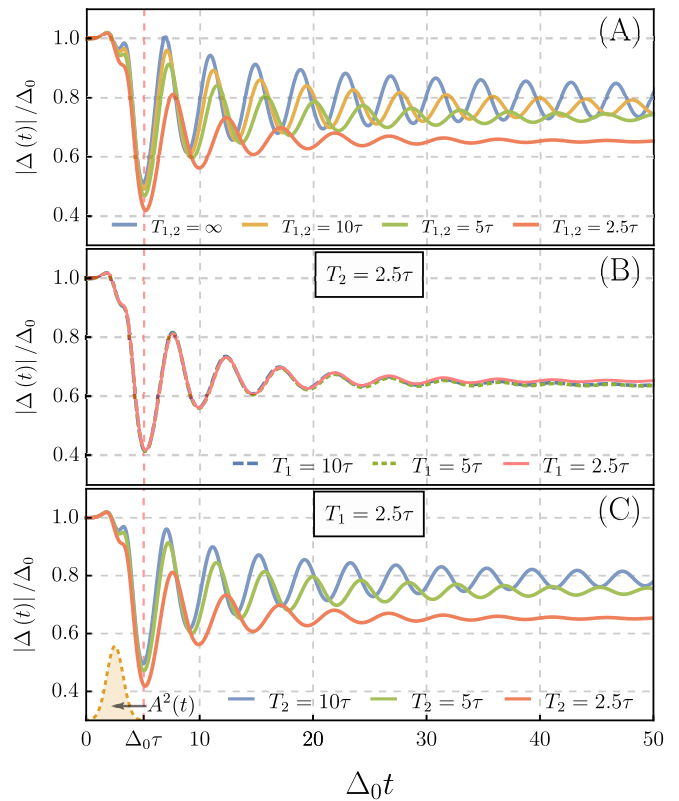} 
\par\end{centering}
\caption{Theoretical results for the time-dependent gap for different values
of the relaxation times $T_{1}$ and $T_{2}$ (in terms of the pump
duration $\tau$). In (A), $T_{1}$ and $T_{2}$ are set to be equal.
Here, $\Delta_{0}$ is the zero temperature, equilibrium value of
the gap function. In (B), $T_{2}$ is fixed at $2.5\tau$ while $T_{1}$
is varied. Panel (C) shows the opposite limit where $T_{2}$ is fixed
at $2.5\tau$ while changing $T_{1}$. In all these panels, a short
Gaussian pulse is considered (also shown in Panel C). \label{fig_theory1}}
\end{figure}

The vector potential $\bfaa(t)$ is related to the electric field
of the pump via $\bfee_{\text{pump}}=-\frac{\partial}{\partial t}\bfaa$.
In our experiment, it takes the form $\bfaa(t)=A_{0}\theta(-t)\theta(\tau-t)\hat{\bfe}_{\text{pump}}e^{-(t-\tau/2)^{2}/2\sigma^{2}}\cos(\omega_{\text{pump}}t)$
with center frequency $\omega_{\text{pump}}$, temporal width $\sigma$,
linear polarization vector $\hat{\bfe}_{\text{pump}}$, and duration
$\tau$. For the calculations in Fig.~\ref{fig_exp}(D) and ~\ref{fig_theory2},
which refer to our experiments on NbN, we consider a long pulse with
$\tau=10\pi/\Delta_{0}$, $\sigma=\tau/5$, $A_{0}=\sqrt{0.75\Delta_{0}}$
and $\omega_{\mathrm{pump}}=1.41\Delta_{0}$, corresponding to a subgap
frequency. To compare with previous experiments involving short pulses,
such as Ref.~\cite{PhysRevLett.111.057002}, in Fig.~\ref{fig_theory1}
we consider a Gaussian-shaped short pulse $\bfaa(t)=A_{0}\theta(-t)\theta(\tau-t)\hat{\bfe}_{\text{pump}}e^{-(t-\tau/2)^{2}/2\sigma^{2}}$
with $\tau=5/\Delta_{0}$, $A_{0}=\sqrt{1.5\Delta_{0}}$.

To describe the gap dynamics, it is convenient to use Anderson pseudospins
$\bfss_{\bfk}=\psi_{\bfk}^{\dag}\frac{\boldsymbol{\sigma}}{2}\psi_{\bfk}$,
with Nambu spinor $\psi_{\bfk}=(c_{\bfk,\uparrow},c_{-\bfk,\downarrow}^{\dag})^{T}$
and Pauli matrices $\boldsymbol{\sigma}$. The Hamiltonian then takes the
simple form $H_{\text{BCS}}=-\sum_{\bfk}\bfbb_{\bfk}\cdot\bfss_{\bfk}+\frac{|\Delta|^{2}}{V_{0}}$,
with a pseudo magnetic field $\bfbb_{\bfk}=2(\Delta',-\Delta'',-\bar{\xi}_{\bfk+e_{0}\bfaa})$,
where $\Delta=\Delta'+i\Delta''$, and $\bar{\xi}_{\bfk+e_{0}\bfaa}=\frac{1}{2}(\varepsilon_{\bfk+e_{0}\bfaa}+\varepsilon_{\bfk-e_{0}\bfaa})-\mu$.
Importantly, the magnetic field depends itself on the state of the
pseudospins via $\Delta=-V_{0}\sum_{\bfk}\av{S_{\bfk}^{-}}$.

In equilibrium, all spins are aligned with the field direction and
their expectation value is given by $\av{\bfss_{\bfk,\text{eq}}}=\frac{1}{2}\hat{\bfs}_{\bfk,\text{eq}}\tanh\Bigl(\frac{E_{\bfk}}{2T_{i}}\Bigr)$.
Here, $T_{i}$ denotes the initial temperature, $E_{\bfk}=\sqrt{|\Delta|^{2}+\xi_{\bfk}^{2}}$
is the Bogoliubov quasiparticle dispersion, and $\hat{\bfs}_{\bfk,\text{eq}}=\bigl(\cos\phi\sin\theta_{\bfk},-\sin\phi\sin\theta_{\bfk},-\cos\theta_{\bfk}\bigr)$
is a unit vector denoting the direction of the pseudospins. The polar
angle is determined by the ratios $\sin\theta_{\bfk}=|\Delta|/(2E{}_{\bfk})$
and $\cos\theta_{\bfk}=\xi_{\bfk}/(2E_{\bfk})$, whereas $\phi$ is
the phase $\Delta=|\Delta|e^{i\phi}$.

The pump pulse $\bfaa(t)$ changes the band dispersion, which in turn
changes the $z$-component of the pseudo magnetic field $B_{\bfk}^{z}$.
Within BCS, the spins precess around the new $\bfbb_{\bfk}$ according
to $\frac{d\av{\bfss_{\bfk}}}{dt}=\av{\bfss_{\bfk}}\times\bfbb_{\bfk}$.
Importantly, the pseudospin dynamics is immediately fed back into
the magnetic field via the gap equation. Due to parity symmetry, only
even-order terms of $\bfaa(t)$ appear \cite{BCS,Anderson-RPA-BCS},
and the oscillation frequency of the gap during the pump is a multiple
of $2\omega_{\text{pump}}$.

By setting $2\omega_{\text{pump}}\approx2\Delta_{0}$, the pump pulse
resonantly drives the coherent $2\Delta$ gap oscillations after the
pump pulse is turned off (i.e. $t>\tau$), similarly to interaction
quenches~\cite{Tsuji}. However, as none of the quench dynamics predicted
by time-dependent solutions of the BCS Hamiltonian (phases I-III described
above) is observed experimentally in NbN (see Fig.~\ref{fig_exp}),
we go beyond this description and include phenomenologically damping
in the pseudospin equations of motion. The microscopic origin of these
terms will be discussed below. In analogy with the general problem
of spin precession, we introduce longitudinal ($T_{1}$) and transverse
$(T_{2})$ relaxation rates:

\begin{align}
\frac{d\av{\bfss_{\bfk}}}{dt}= & \av{\bfss_{\bfk}}\times\bfbb_{\bfk}
-\frac{\av{\bfss_{\bfk}}\cdot\hat{\bfs}_{\parallel,\bfk}^{*}-|\av{\bfss_{\bfk}^{*}}|}{T_{1}}\hat{\bfs}_{\parallel,\bfk}^{*}\nonumber \\
 & -\sum_{i=1}^{2}\frac{\av{\bfss_{\bfk}}\cdot\hat{\bfs}_{\perp,\bfk}^{*,i}}{T_{2}}\hat{\bfs}_{\perp,\bfk}^{*,i}\label{eq:3}
\end{align}

Here, $\av{\bfss_{\bfk}^{*}}[T_{*}(t)]=\frac{1}{2}\hat{\bfs}_{\parallel,\bfk}^{*}[T_{*}(t)]\tanh\bigl(\frac{\sqrt{\xi_{\bfk}^{2}+\Delta_{*}^{2}}}{2T_{*}(t)}\bigr)$
is the thermalized pseudospin configuration at time $t$ at an effective
temperature $T_{*}$. The two vectors $\hat{\bfs}_{\perp,\bfk}^{*,i}$
span the plane perpendicular to the equilibrium pseudospin direction
$\hat{\bfs}_{\parallel,\bfk}^{*}$. Physically, the time scale $T_{1}$
is related to a redistribution of the quasiparticles, whereas the
time scale $T_{2}$ is related to the relaxation of the gap to the
thermalized value $\Delta_{*}$.

To compute $\hat{\bfs}_{\parallel,\bfk}^{*}$ and the effective temperature
$T_{*}$, we consider that all the energy deposited in the electronic
subsystem by the pump is converted into a change in the internal energy
$\mathcal{E}(t)=[\av{H_{\text{BCS}}(t)}_{A=0}-\av{H_{\text{BCS}}}_{i}]$
(see also Ref.~\cite{PhysRevB.95.104507}). Here, the expectation
value is calculated in the time-evolved BCS state according to Eq.~\eqref{eq:3}
and $\av{H_{\text{BCS}}}_{i}$ is the initial ground state energy.
From $\mathcal{E}(t)$, we extract both $T^{*}$ and $\Delta^{*}$,
which are themselves function of time while the pump is turned on.
Once the pump is turned off, energy is no longer deposited in the
electronic subsystem, and thus $T^{*}\left(t>\tau\right)=T^{*}\left(\tau\right)\equiv T_{f}$.
Fig. \ref{fig_theory2}(A) shows $\mathcal{E}(t)$ for different values
of $T_{1}$ and $T_{2}$. The parameters used are the same as in Fig.
\ref{fig_exp}(D). Clearly, the effects of $T_{1}$ and $T_{2}$ kick
in when the pump is weak, as the relaxation processes redistribute
the energy within the electronic subsystem. In Fig. \ref{fig_theory2}(B),
we show how the ``final'' temperature $T^{*}\left(\tau\right)\equiv T_{f}$
depends on the pump fluence. As expected, for sufficiently strong
pumps, the superconducting state can be completely melted by heating.

Using this semi-phenomenological approach, we can capture, as shown
in Fig. \ref{fig_exp}(D), the experimentally-observed gap dynamics
of NbN shown in Fig. \ref{fig_exp}(C). In this calculation, we set
$T_{1}=2T_{2}=1.5\tau$. In contrast to the case with no damping,
$T_{1}=T_{2}=\infty$ (Fig. \ref{fig_exp}(D)), we find that the oscillations
of $|\Delta(t)|$ are quickly suppressed after the pulse is turned
off, and that a continuous and slow increase of $1-|\Delta(t)|/\Delta_{0}$
takes place over the time scale of tens of picoseconds. This characteristic
behavior has also been recently observed in ultraclean samples of
Nb$_{3}$Sn, with a larger post-pump suppression of the gap~\cite{wang_unpublished}.

To correctly capture the experimental observations, it is crucial
to restrict $T_{2}$ to the time scale of the order of the pump duration. To further
elucidate how $T_{1}$ and $T_{2}$ affect the time-evolution of $|\Delta(t)|$,
in Fig. \ref{fig_theory1} we explore different parameter regimes.
In order to highlight the effects of $T_{1}$ and $T_{2}$, and also
to make connection with experiments using short pulses \cite{PhysRevLett.111.057002},
we consider a short Gaussian-shaped pulse of duration $\tau=5/\Delta_{0}$.
As expected, when $T_{1,2}\gg\tau$, the behavior of $\left|\Delta(t)\right|$
is essentially the same as of the system without damping (Fig. \ref{fig_theory1}A).
As $T_{1,2}$ decrease, the damping increases and the gap oscillations become noticeably damped for $T_{1,2}$ of the same order as the pump duration $\tau$. To disentangle the
contributions of $T_{1,2}$, we show $|\Delta(t)|/\Delta_{0}$ for
fixed $T_{2}$ ($T_{1}$) and changing $T_{1}$ ($T_{2}$) in panel
B (C). It is evident that the oscillatory behavior of $|\Delta(t)|$
is much more sensitive on the transverse relaxation $T_{2}$ than
on the longitudinal relaxation $T_{1}$, which only affects weakly
the asymptotic value of the gap. 

We therefore conclude that our experimental observations using long
pulses suggest a dominant $T_{2}$ process in NbN. It is interesting
to note that signatures of damping were also present in previous experiments
on the same material but using short pulses~\cite{PhysRevLett.111.057002}.
Although oscillations were observed in that case after the pump was
off, their decay was reported to be much stronger than the polynomial
$1/\sqrt{t}$ decay predicted by the coherent BCS dynamics. Comparison
with our results in Fig. \ref{fig_theory1}A reveals that this effect
may be explained by the same damping processes revealed in our experiment.

Although $T_{1}$ and $T_{2}$ are phenomenological quantities, it
is important to discuss their possible microscopic origins. As we
explained above, $T_{1,2}$ processes arise within the electronic
subsystem, before equilibration with the lattice. Because the BCS
Hamiltonian is integrable \cite{PhysRevB.72.144524, Yuzbashyan-BCS-Exact-Solution, PhysRevLett.96.097005}, any damping
must arise from non-BCS effects. Residual interactions between the
Bogoliubov quasiparticles, which are neglected in the mean-field BCS
approach, could provide a mechanism for quasiparticle relaxation,
which affects $T_{1}$. Moreover, the Higgs (amplitude) mode excited
resonantly by the laser pump disperses into the quasiparticle continuum
\cite{Benfatto-Nonrelativistic-Higgs,Tsuji}. As a result, one expects
damping of the amplitude mode, which should affect the $T_{2}$ process.

\emph{Conclusions.\textendash{}} In this paper, we established a semi-phenomenological
framework that allows us to incorporate damping in the picosecond
time-evolution of the gap function of an $s$-wave superconductor
subject to an intense THz pulse. In the pseudospin language, damping
arises from a longitudinal process $T_{1}$ (related to quasiparticle
relaxation) and from a transverse process $T_{2}$ (related to relaxation
of the gap). Our experimental results reveal that, in NbN, for large-amplitude
long pump pulses, the picosecond evolution of the gap function is
different than that expected for coherent BCS-like dynamics. Instead,
we showed that the experimental behavior is consistent with a dominant
$T_{2}$ process that arises within the electronic subsystem, and
that has the same time scale as the duration of the pump. Future application
of this approach to different superconductors will allow one to distinguish
the type of relaxation processes dominant in each system. 
\begin{acknowledgments}
We thank M. Schütt for fruitful discussions and N. P. Armitage for
providing the sample. T.C. and R.M.F. are supported by the Office
of Basic Energy Sciences, U.S. Department of Energy, under award DE-SC0012336.
J.W. and X.Y. acknowledge supported by the Army Research office under
award W911NF-15-1-0135 (THz spectroscopy). P.P.O. acknowledges support from Iowa State University Startup Funds.
\end{acknowledgments}

\bibliography{Biblio}

\end{document}